\documentclass[preprint,showpacs,superscriptaddress,aps]{revtex4-1}
\usepackage{amssymb}
\usepackage{amsmath,amssymb,graphicx}
\usepackage{graphicx}
\usepackage{dcolumn}
\usepackage{bm}
\def\be{\begin{equation}}
\def\ee{\end{equation}}
\def\bea{\begin{eqnarray}}
\def\eea{\end{eqnarray}}

\def\lsim{\raise0.3ex\hbox{$\;<$\kern-0.75em\raise-1.1ex\hbox{$\sim\;$}}}
\def\gsim{\raise0.3ex\hbox{$\;>$\kern-0.75em\raise-1.1ex\hbox{$\sim\;$}}}

\newcommand{\mpi}{M_{\pi}}

\newcommand{\GeV}{\,\text{GeV}}
\newcommand{\mk}{M_K}
\newcommand{\mtau}{m_\tau}

\newcommand{\pk}{p_K}
\newcommand{\ppi}{p_\pi}

\newcommand{\diff}{\text{d}}

\newcommand{\Lagr}{\mathcal{L}}

\renewcommand{\Im}{\text{Im}\,}

\usepackage{pstricks}
\begin{document}

\title{Exploring new physics contributions to $CP$ violation in
$\tau^- \to K^-\pi^0\nu_{\tau} $ }

\author{David Delepine}
\email{delepine@fisica.ugto.mx}
\affiliation{{\fontsize{10}{10}\selectfont{Division de Ciencias e
Ingenier\'ias,  Universidad de Guanajuato, C.P. 37150, Le\'on,
Guanajuato, M\'exico.}}}

\author{Gaber Faisel}
\email{gaberfaisel@sdu.edu.tr}

\affiliation{{\fontsize{10}{10}\selectfont{Department of Physics,
Faculty of Arts and Sciences, S\"uleyman Demirel University,
Isparta, Turkey 32260.}}}

\author{ Carlos A. Ramirez}
\email{jpjdramirez@yahoo.com}
\affiliation{{\fontsize{10}{10}\selectfont{Depto. de F\'isica,
Universidad de los Andes, A. A. 4976-12340, Bogot\'a, Colombia.}}}

\begin{center}

\begin{abstract}

A general analysis of possible violation of CP in processes like
$\tau \to K\pi \nu$, for unpolarized $\tau$ is presented. In this
paper, we derive the new contributions to the effective
Hamiltonian governs $\vert\Delta S \vert=1$ semileptonic tau
decays in the framework of two Higgs doublet model with generic
Yukawa structure and Leptoquarks models. Within these models, we
list all operators, in the effective Hamiltonian and provide
analytical expression for their corresponding Wilson coefficients.
Moreover, we  analyze the role of the different contributions,
originating from the scalar, vecor and tensor hadronic currents,
in generating direct CP asymmetry in the decay rate of $\tau^-\to
K^-\pi^0\nu_\tau$. We show that non vanishing direct CP asymmetry
in the decay rate of $\tau^-\to K^-\pi^0\nu_\tau$ can be generated
due to the presence of both, the weak phase in the Wilson
coefficient corresponding to the tensor operator and the strong
phase difference resulting from the interference between the form
factors expressing the matrix elements of the vector and tensor
hadronic currents. After taking into account all relevant
constraints, we find that the generated direct CP asymmetry is of
order $10^{-8}$ which is several orders of magnitude larger than
the standard model prediction. We show also that, in two Higgs
doublet model with generic Yukawa structure , direct local or non
integrated CP violation can be as large as $0.3$ \% not far from
experimental possibilities. This kind of asymmetry can be
generated due to the interference between vector and scalar
contributions with different weak phases which is not the case in
the SM.

\end{abstract}
\end{center}
\pacs{}

\maketitle

\section{Introduction}
Probably the most evident fail of the Standard Model (SM) is the
absence of a mechanism to explain baryogenesis, even if all CP
violation (CPV) processes measured by now are consistent with SM
predictions \cite{cpv,pdg}. At the moment CPV has been observed
only in non leptonic decays of kaons, $B$ and $B_s$ and recently
in $D$ \cite{cpvind}.  In the leptonic sector,  the quantum mixing
of neutrinos yields a source for generating a complex phase
\cite{Fukuda:1998mi,Ahmad:2002jz}. This phase is necessary for
having CPV that can  be measured in $\nu_\mu-\nu_e$ and $\bar
\nu_\mu-\bar \nu_e$ oscillations which are experimentally
accessible with  the Tokai-to-Kamioka (T2K) experiment
\cite{Abe:2013hdq}. Recently, in Ref.\cite{Abe:2019vii}, the T2K
collaboration has reported a measurement that shows an indication
of CPV in the  neutrino sector at $3\sigma$ confidence level.
Particularly, at this confidence level, the reported measurement
are in favor of large enhancement of the neutrino oscillation
probability.  The measurement also excludes values of a complex
phase that can lead to a large enhancement of the observed
anti-neutrino oscillation probability at $3\sigma$ confidence
level. Decays involving leptons like $K_L^0\to \pi^- l^+ \nu_l,\
\pi^+\pi^- e^+e^-$, where CPV has been measured can be understood
as CPV in the meson sector. SM predictions for direct CPV in the
leptonic sector tell us that it should be very small so its
observation would be a clear signal of new physics.

Decays like $\tau\to K\pi \nu$ involve at least two kinds of CPV
contributions: direct CPV and the \lq known' CPV if neutral kaons
are involved. Direct CPV is the same for $\tau^-\to \bar K^0
\pi^-\nu_\tau$, $\tau^-\to K^-\pi^0 \nu_\tau$ and so on, because
the transition $\tau \to s\bar u \nu_\tau$ is the same. Notice
that if additional neutral pions are present the conclusions are
identical.

Earlier searches by CLEO \cite{Bonvicini:2001xz} and Belle
Collaborations \cite{Bischofberger:2011pw} for local or
nonintegrated CPV in the decay $\tau^-\to K_S\pi^-\nu_\tau$ didn't
find any CPV signal (see the corresponding section in this
article). The integrated CPV has been searched  by the BaBar
collaboration with the result

\begin{eqnarray}
A_{\rm CP}^{\rm exp.} ={\Gamma(\tau^+\to K_S\pi^+ \bar \nu_\tau)-
\Gamma(\tau^-\to K_S\pi^- \nu_\tau) \over \Gamma(\tau^+\to
K_S\pi^+ \bar \nu_\tau)+\Gamma(\tau^-\to K_S\pi^- \nu_\tau)
}=-(0.36 \pm 0.23 \pm 0.11)\%
\end{eqnarray}

According to the Standard Model (SM) this process occurs via the
$\tau^- \to s \bar{u}\nu_{\tau}$ transition and no direct CPV
signal is expected. However due to the CPV in mixing in
$K^0-\bar{K}^0$ the total signal should be
\cite{Grossman:2011zk,Bigi:2005ts,Calderon:2007rg}

\begin{eqnarray}
A_{\rm CP}^{\rm theo.} \simeq 2{\rm Re}\ \epsilon_K
=2\cdot(0.166(2))\%= 0.332(2) \%
\end{eqnarray}

There is a 2.8 sigma discrepancy that may indicate the presence of
direct CPV, absent in the SM. However experimental details as the
efficiency in the $K_S$ detection has to be taken into account
properly as was pointed out by Grossman and Nir in
\cite{Grossman:2011zk}. Any real discrepancy is direct CPV and
therefore is new physics (NP) and it should be present in related
channels like $\tau^-\to K^- \pi^0 \nu_\tau$ and so on. Possible
direct angular integrated CP violation in the modes $\tau^-\to
K_S\pi^-\nu_\tau$ and $\tau^-\to K^-\pi^0\nu_\tau$ has been
studied in the literature in Refs.
\cite{Bigi:2005ts,Calderon:2007rg,Devi:2013gya,Cirigliano:2017tqn,Dhargyal:2016kwp,Dhargyal:2016jgo}
and
Refs.\cite{Kuhn:1996dv,Tsai:1996ps,Choi:1998yx,Delepine:2005tw,Delepine:2006fv,Delepine:2007qg,Kimura:2014wsa}
respectively.

In Ref.\cite{Devi:2013gya}, it was shown that angular integrated
direct CPV asymmetry can not be produced by a simple new scalar
interactions, even if they provide new weak phases. As it is well
known, to have direct CPV one needs two interfering contributions
with different weak and strong phases. In these decays once the
angular integral is done the interfering signal vanish. However a
direct CPV signal may remain if the angular integration is partial
or no integration is done (local CPV) at all.

Moreover, as shown in Ref.\cite{Devi:2013gya}, another possibility
is to  have nonstandard tensor interactions with non vanishing
weak phases. In this case the interfering vector and tensor
interactions remain after the full angular integration and a CPV
signal is obtained. However in a recent study, it was shown that
this interference is severely suppressed due to the bounds from
the neutron electric dipole moment and $D-\bar D$ mixing
constrains \cite{Cirigliano:2017tqn}.

An estimation of the direct CPV in $\tau^- \to K^- \pi^0\nu_{\tau}
$, within SM framework showed that the calculated asymmetry is
negligibly small of order $10^{-12}$ \cite{Delepine:2005tw}. This
result motivated further studies of CP violation in this decay
mode within the framework of supersymmetric extension of the SM
\cite{Delepine:2006fv,Delepine:2007qg}. In minimal supersymmetric
extension of the SM with $R$ parity conservation, direct CP
asymmetry of order $O(10^{-7})$ can be generated through the
interference between the vector and tensor interactions
\cite{Delepine:2006fv}. On the other hand, within supersymmetric
extensions of SM with allowed R parity violating terms, no direct
CP asymmetry in the decay rate can be generated at tree-level due
to the absence of tensor interactions \cite{Delepine:2007qg}. In
Refs.\cite{Kuhn:1996dv,Choi:1998yx,Kimura:2014wsa}, it was pointed
out that CP violation in $\tau^-\to K^-\pi^0\nu_\tau$ can arise in
multi Higgs models with complex couplings in the quark sector due
to the interference of the vector and scalar quark currents.

The aim of this paper is to analyze how to generate a $CP$
violation in the semileptonic $\vert\Delta S \vert=1$ tau decays
in the integrated and nonintegrated $CP$ asymmetries. We shall
apply these considerations to specific models of New Physics (NP)
such as 2HDM III and in the SM extensions with scalar leptoquarks.
With the presence of new weak phases and new tensor operator, we
analyze the direct CP violating effects in the decay rate of
$\tau^-\to K^-\pi^0\nu_\tau$.

This paper is organized as follows. In Sec.~\ref{ge}, we present
the effective Hamiltonian describing  semileptonic $\vert\Delta S
\vert=1$ tau decays, $\tau^- \to s \bar{u}\nu_{\tau}$ transition,
in the presence of NP beyond SM. Based on this Hamiltonian, we
derive the general expression of the differential decay width of
the decay process $\tau^-\to K^-\pi^0\nu_\tau$. Switching off NP
contributions to the differential decay width, we show in
Sec.~\ref{icpv} and \ref{lcpv} that no direct CP asymmetry in the
decay rate of $\tau^-\to K^-\pi^0\nu_\tau$ can be generated in the
SM at tree-level.

In Sec.~\ref{Hig}, we derive the analytic expressions of the
Wilson coefficients up to one loop level originating from  the
charged Higgs mediation in 2HDM III. Leptoquarks contributions to
these processes are presented in \ref{Leptoquark}. In
Sec.~\ref{Hig} also, we give our estimation of the direct CP
asymmetry in the decay rate of $\tau^-\to K^-\pi^0\nu_\tau$. In
Sec.~\ref{lcpv} we give our prediction for the local CP violation
in the same decay mode.  Finally, in Sec.~\ref{sec:conclusion}, we
give our conclusion.

\section{Effective Hamiltonian and the differential decay width of
$\vert \Delta S \vert=1$ $\tau $ decays  \label{ge}}

In the presence of NP beyond SM, the effective Hamiltonian governs
$\vert \Delta S \vert=1$ $\tau $ decays transition, taking into
consideration the parity conservation in the $k\to\pi$ matrix
elements, can be expressed as %
\be {\mathcal H}_{eff} =
-\frac{G_F}{\sqrt{2}}V^\star_{us}\sum_{i=V, A,S,P,T}
C_i(\mu)\,Q_i(\mu),\label{Hef}\ee where $V_{us}$ is the $us$
Cabibbo-Kobayashi-Maskawa (CKM) matrix element and $Q_i$ represent
the four-fermion local operators at low energy scale $\mu\simeq
m_\tau$ where

\bea Q_{V} &=&\big(\bar{\nu}_\tau \gamma_\mu\tau\big)\big(\bar{s} \gamma^\mu  u\big) ,\nonumber\\
Q_{A} &=&\big(\bar{\nu}_\tau \gamma_\mu \gamma_5 \tau\big)\big(\bar{s} \gamma^\mu  u\big) ,\nonumber\\
Q_{S} &=& \big(\bar{\nu}_\tau  \tau\big)\big(\bar{s}  u\big) ,\nonumber\\
Q_{P} &=& \big(\bar{\nu}_\tau \gamma_5 \tau\big)\big(\bar{s}  u\big) ,\nonumber\\
Q_{T} &=& \big(\bar{\nu}_\tau \sigma_{\mu \nu } (1+ \gamma_5) \tau
\big)\big(\bar{s}\sigma^{\mu \nu } u\big) ,\label{Qi}\eea with
$\sigma_{\mu\nu}= \frac{i}{2} [\gamma_\mu, \gamma_\nu]$. The
Wilson coefficients, $C_i$, corresponding to the operators $Q_i$
can be expressed as \be C_i = C^{SM}_i+C^{NP}_i, \ee  where
$C^{SM}_i$ and $C^{NP}_i$ represent SM and NP contributions to the
Wilson coefficients respectively. In order to proceed to write the
amplitude we need to calculate the matrix elements of the
operators in the effective Hamiltonian. For this, we need to
assign the momenta of the particles involved in the decay process.
We express the momenta as

\be \tau^-(p_\tau)\to K^- (p_k) + \pi^0 (p_\pi) + \nu_\tau
(p_\nu). \ee The matrix elements of the hadronic currents in the
$Q_i$ operators, in Eq.(\ref{Qi}), are usually parameterized in
terms of particles momenta and form factors. Due to parity
conservation in the $K\to\pi$ matrix elements we need only to
calculate the matrix element of the vector, scalar and tensor
quarks currents only. The matrix element of the vector quark
current can be expressed as
\begin{align}
 \langle K^- \pi^0|\bar s\gamma^\mu u|0\rangle&= \frac{1}{\sqrt{2}}
 \bigg((\pk-\ppi)^\mu f_+(s)+(\pk+\ppi)^\mu f_-(s)\bigg),
\label{vec}\end{align} and \be
f_-(s)=\frac{\Delta^2_{K\pi}}{s}\big(f_0(s)-f_+(s)\big), \ee here
$s$ is the invariant mass defined as $s=(\pk+\ppi)^2$ of the $\pi
K$ system and we have defined $\Delta^2_{K\pi} = \mk^2-\mpi^2$.
The matrix element of the scalar quark current can be obtained
from Eq.(\ref{vec}) by taking the divergence in the usual form and
hence we get \be \langle K^- \pi^0|\bar s u|0\rangle =
\frac{(\mk^2-\mpi^2)} {\sqrt{2}(m_s-m_u)}f_0(s) =
\frac{\Delta^2_{K\pi}}{\sqrt{2}(m_s-m_u)}f_0(s), \label{f2}\ee
$m_{s,u}$ denote $s,u$ current quark masses. Finally, the matrix
element of the tensor quarks current, $\langle K^- \pi^0|\bar
s\sigma^{\mu\nu} u|0\rangle$, can be expressed as
\cite{Cirigliano:2017tqn}
\begin{align}
\langle K^- \pi^0|\bar s\sigma^{\mu\nu}
u|0\rangle&=\frac{i(\pk^\mu\ppi^\nu-\pk^\nu\ppi^\mu)}{\sqrt{2}\mk}B_T(s).\label{f3}
\end{align}
For specific models, like the ones under investigation in this
study, the  Wilson coefficients $C_i$  can be expressed in terms
of only three independent coefficients. Particularly, in these
models, we have $C_A = - C_V$ and $C_P = C_S $ and hence we are
left with only three independent Wilson coefficients, namely
$C_V$, $C_S $ and $C_T$. This indicates that, the set of the
operators in Eq.(\ref{Qi}), within these models, can be rewritten
in terms of just three independent operators as
$\big(\bar{\nu}_\tau \gamma_\mu L \tau\big)\big(\bar{s} \gamma^\mu
u\big)$, $ \big(\bar{\nu}_\tau R \tau\big)\big(\bar{s} u\big)$ and
$ \big(\bar{\nu}_\tau \sigma_{\mu \nu } R \tau
\big)\big(\bar{s}\sigma^{\mu \nu } u\big)$ where $L,R=1\mp
\gamma_5$. The total amplitude, $\mathcal{A}$, of
$\tau^- \rightarrow K^-\pi^0 \nu_\tau$ decay can be expressed as%
\begin{eqnarray}
\mathcal{A} &=&-\frac{G_{F}V_{us} C_V}{\sqrt{%
2}}\bigg\{ \bigg((\pk-\ppi)^\mu f_+(s)+(\pk+\ppi)^\mu f_-(s)\bigg)
\Big(\bar{u}(p_\nu)\gamma _{\mu }Lu(p_\tau)\Big)\nonumber
\\&+&
\frac{C_S\,\Delta^2_{K\pi}}{(m_s-m_u)\,
C_V}f_0(s)\Big(\bar{u}(p_\nu)R u(p_\tau)\Big)\nonumber
\\&+& i\frac{(\pk^\mu\ppi^\nu-\pk^\nu\ppi^\mu)\,
C_{T}}{\mk C_V}B_T(s)\Big(\bar{u}(p_\nu)\sigma _{\mu\nu } R
u(p_\tau) \Big)\bigg\} \ .\nonumber
\end{eqnarray}%

The differential decay width  is given as
\begin{eqnarray}
\frac{\diff\Gamma}{\diff s}&=& G_F^2|V_{us}|^2 |C_V|^2 S_\text{EW}
\frac{\lambda^{1/2}(s,\mpi^2,\mk^2)(\mtau^2-s)^2 \Delta^4_{K\pi} }{1024\pi^3\mtau s^3}\nonumber\\
&\times&\bigg[\frac{(\mtau^2+2s)\lambda(s,\mpi^2,\mk^2)}{3\mtau^2
\Delta^4_{K\pi} }\bigg(|f_+(s)
-T(s)|^2+\frac{2(\mtau^2-s)^2}{9s\mtau^2}|T(s)|^2\bigg)+|S(s)|^2\bigg],
\label{decay_width} \end{eqnarray} where $\lambda(x,y,z)$ is given
by $\lambda(x,y,z) = x^2+y^2 +z^2 -2 xy -2 xz -2 yz$,
$S_\text{EW}=1.0194$~\cite{Marciano:1985pd,Marciano:1988vm,Braaten:1990ef}
accounts for the electroweak running down to $\mtau$ and
\begin{eqnarray}
S(s)&=& f_0(s)\bigg(1+\frac{s\, C_S}{\mtau(m_s-m_u)\, C_V} \bigg),\nonumber\\
T(s)&=&\frac{3s}{\mtau^2+2s}\frac{\mtau \, C_{T}}{\mk C_V} \,
B_T(s). \label{VT1}\end{eqnarray} It should be noted that
Eq.(\ref{decay_width}) above and Eq.(12) of
Ref.\cite{Cirigliano:2017tqn} are inconsistent only by a factor
$1/2$ due to the presence of extra factor $1/\sqrt{2}$ in the
hadronic matrix elements, Eqs.(\ref{vec},\ref{f2},\ref{f3}),
compared to their corresponding ones listed in Eq.(10) in
Ref.\cite{Cirigliano:2017tqn}. This is due to the difference of
the final states $K^-\pi^0$ and $\bar K^0\pi^-$ in our work and in
Ref.[11] respectively. In the isospin symmetry limit, the form
factors of $\tau^-\to K^-\pi^0\nu_\tau$ are not equal to their
corresponding ones in the decay mode $\tau^-\to \bar
K^0\pi^-\nu_\tau$ \cite{Finkemeier:1996dh}. Rather, they are
related by a simple Clebsch-Gordan factor $1/\sqrt{2}$ as given in
Eqs.(4,8) in Ref.\cite{Finkemeier:1996dh}. These parameterizations
had been adopted in our previous studies in
Refs.\cite{GodinaNava:1995jb,Delepine:2006fv}.

 Non vanishing direct CP asymmetry in the decay rate requires the
presence of two types of phases, the weak CP violating phases and
the strong CP conserving phases. The weak CP violating phases can
be generated in the Wilson coefficients upon existence of complex
couplings. On the other hand, the strong CP conserving phases
originate from the phases in the form factors expressing the
matrix elements of the hadronic currents.

Breit-Wigner forms are used to parameterize the contributions of
the different resonances dominating the scalar and vector hadronic
currents. As a consequence, form factors originating from these
currents can be expressed as a summation of Briet-Wigner forms.
Previous studies of CP asymmetries in $\tau\to K \pi \nu_\tau$
decays, for instances Refs.\cite{Delepine:2006fv,Devi:2013gya},
adopted the assumption that the form factor $B_T(s)$ has no strong
phases.  However, this assumption is incorrect as argued in
Ref.\cite{Cirigliano:2017tqn}. As shown in
Refs.~\cite{Ecker:1988te,Ecker:1989yg},  spin-$1$ resonances can
be described equivalently by vector or antisymmetric tensor fields
. Hence the same resonances $K^*(892)$ and the $K^*(1410)$ that
dominate the Briet-Wigner forms in $f_+(s)$ will appear in
$B_T(s)$ as well \cite{Cirigliano:2017tqn}. It should be noted
that, this conclusion can be derived by analyzing the unitarity
relation for the form factors as shown in details in
Ref.\cite{Cirigliano:2017tqn}. Thus, we conclude that $B_T(s)$ has
a strong phase that should be taken into account in the
calculations of the CP asymmetry.

In the SM, at tree-level, the Wilson coefficients $C_i$ reduces to
\be C^{SM}_{V} = -C^{SM}_{A} =
1,\,\,\,\,\,\,\,\,\,\,\,\,\,C^{SM}_{S,P,T} =0,\label{SMWC}\ee This
accounts for the fact that $\tau ^{-}\rightarrow K^{-}\pi
^{0}\nu_{\tau }$, at tree-level, can be generated as a result of
exchanging single $W^-$ boson which contributes only to $Q_V$
operator. Consequently, the quantities $S(s)$ and $T(s)$ in
Eq.(\ref{VT1}) reduce to \be S(s) \to f_0(s),\,\,\,\,\,\,\, T(s)
\to 0,\ee and upon substation in Eq.(\ref{decay_width}) we get
\begin{align}
\label{decay_widthsm} \frac{\diff\Gamma}{\diff
s}\bigg|_{\text{SM}}&=G_F^2|V_{us}|^2 S_\text{EW}
\frac{\lambda^{1/2}(s,\mpi^2,\mk^2)(\mtau^2-s)^2 \Delta^4_{K\pi}}{1024\pi^3\mtau s^3}\notag\\
&\times\bigg[\frac{(\mtau^2+2s)\lambda(s,\mpi^2,\mk^2)}{3\mtau^2
\Delta^4_{K\pi}}|f_+(s)|^2 +|f_0(s)|^2\bigg],
\end{align}
The decay rate of the process $\tau \to K^- \pi^0 \nu_{\tau}$ in
the SM, $\Gamma_{SM}$, can be then obtained upon integrating the
previous equation with respect to the kinematic variable $s$. Thus
we get
\begin{align}
\label{decay_widthsm}
&\Gamma_{SM}=\frac{G_F^2|V_{us}|^2  S_\text{EW} \Delta^4_{K\pi} }{1024\pi^3\mtau}\notag\\
& \int^{m^2_\tau}_{(\mk+\mpi)^2} ds
\bigg(\frac{\lambda^{1/2}(s,\mpi^2,\mk^2)(\mtau^2-s)^2}{ s^3}
\times\bigg[\frac{(\mtau^2+2s)\lambda(s,\mpi^2,\mk^2)}{3\mtau^2
\Delta^4_{K\pi} }|f_+(s)|^2 +|f_0(s)|^2\bigg]\bigg),
\end{align}
The CP asymmetry in total decay rate of $\tau ^{-}\rightarrow
K^{-}\pi ^{0}\nu_{\tau }$ is given by:%
\begin{eqnarray}
A_{CP} &=&\frac{\Gamma (\tau ^{-}\rightarrow K^{-}\pi ^{0}\nu
_{\tau })-\Gamma (\tau ^{+}\rightarrow K^{+}\pi ^{0}\nu _{\tau
})}{\Gamma (\tau ^{-}\rightarrow K^{-}\pi ^{0}\nu _{\tau })+\Gamma
(\tau ^{+}\rightarrow K^{+}\pi ^{0}\nu _{\tau })}.
\end{eqnarray}%
Clearly, from the expression of $\Gamma_{SM}$, direct CP asymmetry
in the decay rate will vanish due to the absence of the weak
phase, $C^{SM}_V$ is real, and also due to the remark that the
form factors $f_+(s)$ and $f_0(s)$ do not interfere and hence the
relative strong phase essential for CP asymmetry vanishes. Thus,
to generate no vanishing CP asymmetry in the decay rate within SM,
it is essential to consider higher order terms contributing to the
amplitude as done in Ref.\cite{Delepine:2005tw}. These terms can
be generated from diagrams with exchanging two $W$ bosons.  Thus,
the generated asymmetry is expected to be very small. As shown in
Ref.\cite{Delepine:2005tw}, the resulting CP asymmetry is
suppressed by the CKM factor $V_{td} \simeq 10^{-3}$ and also by a
higher order suppresion factor $g^2/4\pi M^2_W \simeq10^{-8}$. As
a consequence, the resulting CP rate asymmetry is expected to be
negligible. In fact the asymmetry is of order $10^{-12}$ as
estimated in Ref.\cite{Delepine:2005tw}.

Integrated direct or local CP violation has been discussed above.
However having somehow the ability to measure angular distribution
one can extract the local CPV signal \cite{Kimura:2014wsa} and
$|f_0|$. The CPV signal is produced due to the interference
between the vector and the scalar contributions, as long as they
contribute with different weak phases. Additional CPV observables
are provided by triple products, but in that case polarized taus
are needed  \cite{Kuhn:1996dv} \cite{Choi:1998yx}
\cite{Tsai:1996ps}. Without considering the exotic tensor
interactions, one finds that the local distribution is given as

\begin{eqnarray}
{{\rm d}^2\Gamma^- \over  {\rm d}s  {\rm d}x} &=& { G^2_F V^2_{us}
m_\tau  \over 64\pi^3 s^{3/2}} |{\bf p}_K| |{\bf q}|^2 \left\{
\left[{s\over m_\tau^2} +\left(1-{s\over m_\tau^2}\right) x^2
\right] \left|{\bf p}_K C_V f_+\right|^2 + {\Delta_{K\pi}^4\over
4s} \left|C_Sf_0\right|^2\right.
\nonumber \\
&& \left. +{\Delta_{K\pi}^2\over \sqrt{s}}|{\bf p}_K|{\rm
Re}\left[C_VC_S^*f_+f_0^* \right]x \right\},
\end{eqnarray}
where the negative sign in $\Gamma^-$  corresponds to the QED
charge of $\tau^-$ lepton, $x= \cos\theta$ and $|{\bf
q}|=m_\tau(1-s/m_\tau^2)/2$ \, is the momentum of the $K-\pi$
system (and of the neutrino) in the $\tau$ center of mass.
Similarly $|{\bf p}_K|=\lambda^{1/2}(s,\ m_K^2,\
m_\pi^2)/2\sqrt{s}$ is the kaon (and of the pion) momentum in the
$K-\pi$ center of mass. The angle $\theta$ is the scattering angle
of kaon with respect to the incoming $\tau$, or equivalently the
angle between $\overrightarrow{p}_\pi$ and
$\overrightarrow{p}_\nu$, in the hadronic rest frame. The Local
CPV is defined as

\begin{eqnarray}
&&A_{\rm CP}(s,\ x)  = { {\rm d^2}\Gamma^+/{\rm d}s  {\rm d}x -
{\rm d^2}\Gamma^-/{\rm d}s  {\rm d}x \over {\rm d^2}\Gamma^+/{\rm
d}s {\rm d}x +{\rm d^2}\Gamma^-/{\rm d}s {\rm d}x }\nonumber
\\
&=& { 4m_\tau^2\Delta_{K\pi}^2 \sqrt{s}\ |{\bf p}_K| x\ {\rm
Im}[C_VC_S^*]\ {\rm Im}(f_+ (s) f_0^*(s)) \over
    4s\left(s+ 2m_\tau|{\bf q}|x^2 \right)\left|{\bf p}_K C_V f_+(s)\right|^2  +m_\tau^2\Delta_{K\pi}^4 \left|C_S f_0(s)\right|^2
+4 m_\tau^2 \Delta_{K\pi}^2\sqrt{s}\ |{\bf p}_K| x \ {\rm Re}(C_V
C_S^*) \ {\rm Re}(f_+(s)f_0^*(s)) },   \nonumber\\
\label{LCPV}\end{eqnarray} Similarly to the sign convention in
$\Gamma^-$, the positive sign in $\Gamma^+$ corresponds to the QED
charge of $\tau^+$. Finally, it should be remarked that the local
CPV, defined in Eq,(\ref{LCPV}), is different than the
forward-backward asymmetry, ${\mathcal A}_{FB}(s)$, that can be
obtained upon integrating ${{\rm d}^2\Gamma \over {\rm d}s  {\rm
d}x}$  over $x$. Explicitly, it is defined as
\cite{Rendon:2019awg}

\be{\mathcal A}_{FB}(s)=\frac{\int^1_0 {\rm d}x \frac{{\rm
d}^2\Gamma^- }{{\rm d}s {\rm d}x} -\int^0_{-1} {\rm d}x \frac{{\rm
d}^2\Gamma^- }{{\rm d}s {\rm d}x}  }{\int^1_0 {\rm d}x \frac{{\rm
d}^2\Gamma^- }{{\rm d}s {\rm d}x} +\int^0_{-1} {\rm d}x \frac{{\rm
d}^2\Gamma^- }{{\rm d}s {\rm d}x} } \ee

It is clear that within SM at tree-level, from Eq.(\ref{LCPV}),
the local CPV is vanishing due to the remark that  $C^{SM}_S=0$.
This is not the case of the forward-backward asymmetry ${\mathcal
A}_{FB}(s)$ that has non-vanishing term independent of $C^{SM}_S$
and $C^{SM}_T$ \cite{Rendon:2019awg}. Thus, non-vanishing values
of the local CPV can be used as a probe of physics beyond the SM
as we will discuss in Sec.\ref{lcpv}.

\section{Integrated CP violation asymmetries \label{icpv}}

 We investigate now the direct CP asymmetry for a general new physics
 model that can contribute to the effective Hamiltonian in
 Eq.(\ref{Hef}) with Wilson coefficients denoted by $C^{NP}_i$.
As can be seen from Eqs.(\ref{decay_width},\ref{VT1}), the
hadronic form factor $f_0(s)$, in $S(s)$, does not interfere with
the form factor $f_+(s)$. The absence of this interference leads
to the absence of the strong phase difference between their
contributions to the decay rate. This phase difference is
essential for generating non vanishing direct CP asymmetry. As a
consequence, and for having non-vanishing direct CP asymmetry in
the decay rate, we are left only with the interference between
$B_T(s)$, in $T(s)$, and $f_0(s)$ as a possible source for the
required strong phase difference. However, this interference was
estimated to be small due to Watson's final-state-interaction
theorem\cite{Cirigliano:2017tqn,Watson:1954uc}. Assuming that $NP$
contributions are obtained via integrating out heavy particles,
above the electroweak breaking scale, we can set
 $C^{NP}_V=0$. In this case,  the direct CP asymmetry in total decay rate of
$\tau ^{-}\rightarrow
K^{-}\pi ^{0}\nu_{\tau }$  is given by:%
\begin{eqnarray}
A^{NP}_{CP} &=& -\frac{ G_F^2|V_{us}|^2 S_\text{EW}\,
Im\,C^{NP}_T}{512\pi^3\mtau^2\mk\Gamma_\tau\text{BR}(\tau\to
K\pi\nu_\tau)}
\nonumber \\
 &&\hspace{-1cm}\times\int_{(\mpi+\mk)^2}^{\mtau^2}\diff s
 \frac{\lambda^{3/2}(s,\mpi^2,\mk^2)(\mtau^2-s)^2}{ s^2} |f_+(s)||B_T(s)|
 \sin\big(\delta_+(s)-\delta_T(s)\big),
\label{acpas1}\end{eqnarray} where $\delta_+(s)$, $\delta_T(s)$
are the phases of $f_+(s)$ and $B_T(s)$.  Estimation of the $CP$
asymmetry in the previous equation requires information on the
form factor $f_+(s)$ and $B_T(s)$. Empirically, $\tau\to\pi
K_S\nu_\tau$ spectrum can give information on the form factor
$f_+(s)$ which is mainly dominated by the $K^*(892)$ resonance
\cite{Epifanov:2007rf}. As discussed in
Refs.\cite{Ecker:1988te,Ecker:1989yg,Cirigliano:2017tqn}, spin-$1$
resonances can be described equivalently by vector or
antisymmetric tensor fields. Thus, both  of $f_+(s)$ and $B_T(s)$
can receive contributions from the  same resonances, mainly
$K^*(892)$ resonance. Considering the moduli of the form factors
$f_+(s)$ and $B_T(s)$, inelastic corrections turns to be
negligible and the elastic solution of the unitarity relation for
the form factors can be used \cite{Cirigliano:2017tqn}
 \be f_+(s)=f_+(0)\Omega(s),\qquad B_T(s)=B_T(0)\Omega(s),\label{ffa}
\ee in terms of the Omn\`es factor~\cite{Omnes:1958hv} \be
\label{Omnes} \Omega(s)=\exp\bigg\{\frac{s}{\pi}\int_{s_{\pi
K}}^\infty\frac{\delta(s')}{s'(s'-s)}\bigg\}. \ee In the previous
equation, the phase shift $\delta(s)$ can be approximated by a BW
phase \cite{Cirigliano:2017tqn}  with parameters as determined
in~\cite{Epifanov:2007rf}. As shown in
Ref.\cite{Cirigliano:2017tqn}, the modulus of $f_+(s)$, given in
Eq.(\ref{ffa}), can reproduce well the experimental fit below the
$K^*(1410)$ resonance done in Ref.\cite{Epifanov:2007rf}.

 We turn now to the phase of the form factor $f_+(s)$, namely
$\delta_+(s)$.  Experiments are only sensitive to the modulus of
the form factor. Thus $\delta_+(s)$ cannot be directly estimated
from experiments. Rather, its extraction can be done with the help
of a fit function that preserves the analytic structure of the
form factor \cite{Cirigliano:2017tqn}. As a consequence, the fit
function used in~\cite{Epifanov:2007rf} cannot be helpful for
extracting $\delta_+(s)$. However, as argued in
Ref.\cite{Cirigliano:2017tqn}, still this fit can be useful in
extracting $\delta_+(s)$. This can be achieved through comparing
the phase from the experimental fit~\cite{Epifanov:2007rf} with
$\delta_+(s)$ computed using a BW approximation for the
$K^*(892)$. To account for the inelastic contribution
$\delta^\text{inel}_+(s)$ to $\delta_+(s)$, the authors of
Ref.\cite{Cirigliano:2017tqn} have added the BW phase for
$K^*(1410)\to K^*(892)\pi$ with a coefficient that allows for a
similar phase motion in the vicinity of the $K^*(1410)$. This
leads to the band shown in Fig.~2, in
Ref.\cite{Cirigliano:2017tqn}, which represents a simple
estimation of inelastic effects. The result is consistent with
more refined estimates along the lines
of~\cite{Moussallam:2007qc,Boito:2008fq,Boito:2010me,Bernard:2011ae,Antonelli:2013usa}.
Up on the assumption that the inelastic contributions in
$\delta_T(s)$ are of similar size with an opposite sign, one can
take $\delta_+(s) - \delta_T(s) \sim 2 \delta_+^\text{inel}(s)$
\cite{Cirigliano:2017tqn}.

With the estimation of the form factors following the discussion
above and using $\text{BR}(\tau\to K^- \pi^0\nu_\tau)= (4.33\pm
0.15)\times 10^{-3}$ \cite{Patrignani:2016xqp},
$f_+(0)|V_{us}|=0.2165(4)$~\cite{Patrignani:2016xqp},
$B_T(0)/f_+(0)=0.676(27)$ from lattice QCD~\cite{Baum:2011rm},
particle masses and couplings from~\cite{Patrignani:2016xqp}, we
find that the $CP$ asymmetry can be estimated as \be
\big|A^{NP}_{CP}\big|\lesssim 1.4\times 10^{-2}\, Im \, C^{NP}_T,
\label{ACP}\ee

An investigation of the limits on $\Im C_T$ for  general NP models
follow from electric dipole moment (EDM) of the neutron and
$D$--$\bar{D}$ mixing was carried in
Ref.\cite{Cirigliano:2017tqn}. In the following we present the
derivation of such limits as carried out in Ref
\cite{Cirigliano:2017tqn}. At an energy scale $\Lambda \gg v$, the
decay processes $\tau^- \to K^- \pi^0 \nu_\tau$ receive
contribution from a tensor operator. The operator is originating
from the $SU(3) \times SU(2) \times U(1)$ gauge-invariant
Lagrangian that can be expressed as \cite{Cirigliano:2017tqn} \be
\label{eq:LT0} \Lagr_T = C_{abcd} \ \bar{L}_{La}^i \sigma_{\mu
\nu}  e_{Rb} \, \epsilon^{ij}  \, \bar{q}_{L c}^j \sigma^{\mu \nu}
u_{Rd}  \ + \ \text{h.c.}, \ee  here $a,b,c,d$ are generation
indices and $i,j$ are $SU(2)_L$ indices. In the preceding equation
$q_L$ and $L_L$ stand for the quark and lepton  $SU(2)_L$ doublets
while  $u_R$ and $e_R$  represent the charged up-quark and lepton
$SU(2)_L$ singlets respectively. The tensor operator $Q_T$ listed
in Eq.(\ref{Qi}) is generated from \cite{Cirigliano:2017tqn}
\begin{align}
\label{eq:LT1} \Lagr_T  = C_{3321}\Big[ (\bar{\nu}_{\tau}
\sigma_{\mu \nu}R  \tau)(\bar{s}  \sigma^{\mu \nu} R u) - V_{us}
(\bar{\tau} \sigma_{\mu \nu}R  \tau)(\bar{u} \sigma^{\mu \nu} R u)
\Big]   \ + \ \text{h.c.},
\end{align}
where $R=(1+\gamma_5)/2$ and other terms that involve the charm
and top quark have been omitted. Comparing Eq.(\ref{Qi}) and
Eq.(\ref{eq:LT1}) one finds that  \be \label{C3321} C_{3321} = -
\sqrt{2} G_F V_{us} C_T. \ee The renormalization group
evolution~\cite{Jenkins:2013wua} of the operator $(\bar{\tau}
\sigma_{\mu \nu}R  \tau)(\bar{u} \sigma^{\mu \nu} R u)$, second
term in the square bracket in Eq.(\ref{eq:LT1}), can produce via
insertion an up-quark EDM $d_u (\mu)$  \cite{Cirigliano:2017tqn}
 \be \Lagr_\text{D}=- \frac{i}{2}  d_u (\mu) \bar{u} \sigma^{\mu
\nu} \gamma_5   u F_{\mu \nu}, \ee

Upon solving the RG
following ~\cite{Bellucci:1981bs,Buchalla:1989we,Cirigliano:2017azj}
one obtains  \cite{Cirigliano:2017tqn}
\begin{align}
d_u (\mu) &=   \frac{e \mtau}{v^2}    \frac{V_{us}^2}{\pi^2}   \
\Im C_T (\mu)   \ \log \frac{\Lambda}{\mu}.
\end{align}
The last equation above together with the $90\%$ C.L.\ bound $d_n
= g_T^{u} (\mu) d_u (\mu) < 2.9 \times
10^{-26}\,e\,\text{cm}$~\cite{Baker:2006ts,Afach:2015sja} can be
used to obtain a strong bound on $\Im C_T$. Thus, for a value
$\Lambda\gtrsim 100\GeV$ and using   and the recent lattice
result~\cite{Bhattacharya:2015esa} $g_T^{u} (\mu = \mu_\tau =
2\GeV)  = - 0.233(28)$   one finds the  constraint
\cite{Cirigliano:2017tqn}   \be | \Im C_T (\mu_\tau) | \lesssim
10^{-5}, \label{ctc} \ee It should be noted that, the above
constraint is based on the assumption that there are no other
contributions to $d_n$ that can cancel the effect of $C_T$.

Applying the derived constraint on $\Im C_T(\mu_\tau)$, we get a
model independent prediction  of the direct $CP$ asymmetry as \be
\big|A^{NP}_{CP}\big|\lesssim {\mathcal O} (10^{-7}).
\label{cT2}\ee Clearly, from the above discussion, NP
contributions to the phases of $C^{NP}_5$ are needed to have
non-vanishing direct $CP$ asymmetry. These contributions will be
evaluated below for several NP models.

\subsection{ CP asymmetry of $\tau ^{-}\rightarrow K^{-}\pi
^{0}\nu_{\tau }$  in 2HDM III.}\label{Hig}

\begin{figure}[tbhp]
\includegraphics[width=5cm,height=4cm]{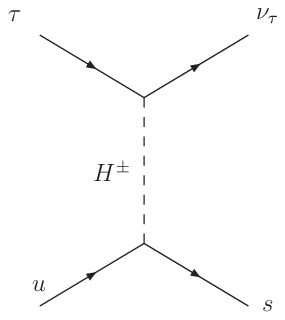}
\hspace{0.2cm}
\includegraphics*[width=5cm,height=4cm]{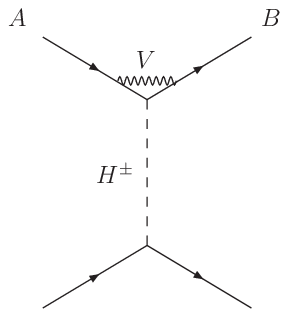}
\hspace{0.2cm}
\includegraphics*[width=5cm,height=4cm]{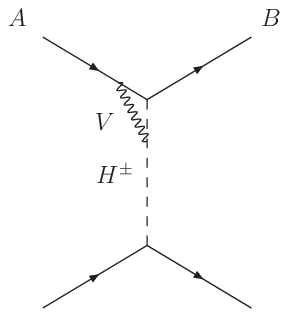}
\hspace{0.2cm}
\includegraphics*[width=5cm,height=4cm]{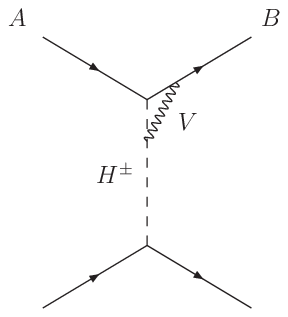}
\hspace{0.2cm}
\includegraphics*[width=5cm,height=4cm]{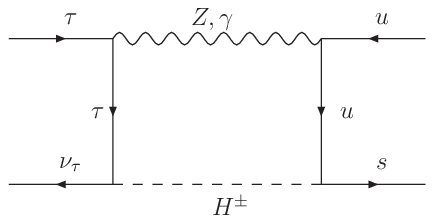}
\hspace{0.2 cm}
\includegraphics*[width=5cm,height=4cm]{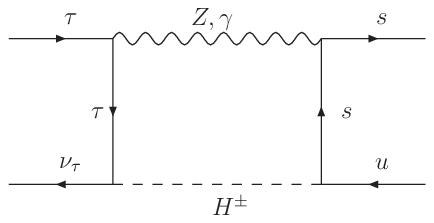}
\hspace{0.2 cm}
\includegraphics[width=5cm,height=4cm]{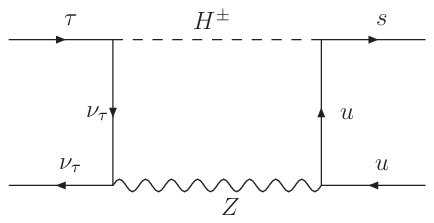}
\hspace{0.2cm}
\includegraphics*[width=5cm,height=4cm]{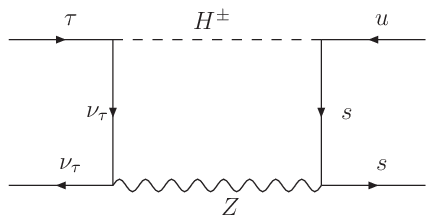}
\medskip
\caption{ Diagrams contributing to the effective Hamiltonian,
${\mathcal H}_{eff}$, in Eq.(\ref{Hef}) up to one loop-level due
to charged Higgs mediation. In the figures, $V$ can be $Z$ boson
or photon or both of them depending on the fermions $A$ and $B$.
In the case that $A$ and $B$ are $\tau$ and $\nu_\tau$, the other
external lines, in the figures, to be understood as representing
up and strange quarks. On the other hand when $A$ and $B$ are up
and strange quarks, the other external lines in the figures are
simply representing $\tau$ and $\nu_\tau$. \label{ZBs}}
\end{figure}

Two Higgs doublet models(2HDM) are simple extensions of the SM in
which the scalar sector of the SM is enlarged to contain new
scalars
\cite{Haber:1978jt,Abbott:1979dt,Arbey:2017gmh,Zarikas:1995qb,Lahanas:1998wf,Aliferis:2014ofa}.
Based on the couplings of the new scalars to quarks and leptons we
can classify these models to several types such as type I, II or
III \cite{Branco:2011iw}. In the two Higgs doublet model with
generic Yukawa structure or simply type III (2HDM III), the
couplings of the new scalars to quarks and leptons can be complex
\cite{Crivellin:2010er,Crivellin:2012ye,Crivellin:2013wna}. As a
consequence, these couplings can serve as the source of the weak
CP violating phases essential for generating non vanishing CP
asymmetries. The effect of these new weak phases on the CP
asymmetry in D meson sector have been investigated in
Refs.\cite{Delepine:2012xw,Delepine:2014mga,Delepine:2017oor}. The
resultant direct CP asymmetry in this model can be enhanced
several orders of magnitudes larger than SM predictions. For
instances, a large direct CP asymmetry  of order $10^{-3}$ can be
achieved in the decay mode $D^0 \to K^+ \pi^- $ after taking into
account all constraints on the parameter space of the model
\cite{Delepine:2017oor}. This value is $6$ orders of magnitude
larger than the standard model prediction. The 2HDM III is also
motivated by its ability to explain some anomalies in $B$ decays.
As shown in Ref.\cite{Crivellin:2013wna}, a 2HDM of type II (like
the MSSM at tree-level) cannot explain the deviations from the SM
in tauonic $B$ decays, while 2HDM III  can account for $B\to
D\tau\nu$ and $B\to D^*\tau\nu$ simultaneously~
\cite{Crivellin:2012ye,Crivellin:2015hha}.

The scalar sector of the 2HDM III consists of two Higgs doublets.
The mass eigenstates constructed from these doublets are $H_0$
(heavy CP-even Higgs), $h_0$ (light CP-even Higgs) and $A_0$
(CP-odd Higgs) and $H^{\pm}$.  In 2HDM III, the charged Higgs
couplings to quarks and leptons can be expressed as
\cite{Crivellin:2010er,Crivellin:2012ye}
\begin{equation}
\mathcal{L}^{eff}_{H^\pm} = \bar{u} {\Gamma_{u s
}^{H^\pm\,LR\,\rm{eff} } }P_R s
+ \bar{u} {\Gamma_{u s }^{H^\pm\,RL\,\rm{eff} } }P_L s+\frac{m_\tau\tan\beta}{v} (\bar\nu_\tau P_R\tau)+h.c.\, ,\\
 \label{Higgs-vertex}
\end{equation}
where \bea {\Gamma_{u s }^{H^\pm\,LR\,\rm{eff} } } &=&
 {\sin\beta \left( V_{12}\frac{m_{s }}{v_d} -
 \sum\limits_{j = 1}^3 \, V_{1j} \epsilon^{ d}_{j2}\tan\beta \right), }
\nonumber\\
{\Gamma_{u s }^{H^ \pm\,RL\,\rm{eff} } } &=& \cos\beta\, \left(
\frac{m_{u }}{v_u} V_{12}-\sum\limits_{j = 1}^3
 V_{j2} \epsilon^{ u\star}_{j1}\tan\beta \right),
 \label{Higgsv}
\eea Here $v_u$ and $v_d$ stand for the vacuum expectations values
of the neutral component of the  Higgs doublets,  tan $\beta =
v_u/v_d$ and finally $V$ is the CKM matrix. Using the
Feynman-rules given in Eq.(\ref{Higgs-vertex}), we can proceed to
derive the relevant Wilson coefficients to our investigation.
Working in the unitary gauge and after integrating out the charged
Higgs in the diagrams in Fig.\ref{ZBs}, we find that the relevant
Wilson coefficients at $\mu= m_H$ scale can be expressed as
\begin{eqnarray}
C^{H^\pm}_{V }&=& -\frac { m^2_W m_s m^2_\tau \tan\beta \sin\beta
}{2 \pi^2 m^4_H\,  v\, V^\star_{us}\,c^2_w} \bigg((-\frac{1}{2}+
s^2_w)(-\frac{1}{2}+\frac{1}{3}s^2_w) \, g(x_s,x_\tau,x_Z) + \frac
{ 4 v^2\,\pi \alpha \, m_u c^2_w }{ 3 m^2_W m_s } k(x_u,x_\tau)
\bigg)\nonumber\\
&\times& \bigg( \frac{m_s V^\star_{us}}{v_d} - \sum\limits_{j =
1}^3 V^\star_{1j} \epsilon^{ d \star}_{j2}\tan\beta
\bigg)\nonumber\\
&-&\frac { 8 m^2_W m_s m^2_\tau \sin\beta s^2_w}{3 v\,
V^\star_{us}\,c^2_w} \bigg((-\frac{1}{2}+ s^2_w)\,
g(x_s,x_\tau,x_Z) + \frac { 4 v^2\,\pi \alpha \, m_u c^2_w }{
m^2_W m_s s^2_w } k(x_u,x_\tau)
\bigg)\nonumber\\
&\times& \bigg(\frac{m_u V^\star_{us} }{v_u} - \sum\limits_{j =
1}^3 V^\star_{j2} \epsilon^{ u}_{j1}\tan\beta \bigg),\nonumber\\
C^{H^\pm}_S &=& \frac{ v\, m_\tau \sin\beta}{ 4\pi^2 m^2_H
V^\star_{us}}\bigg(16 \pi^2 + \frac { 8\,\pi \alpha }{3 }
f(x_u,x_\tau) +\frac { 2 m^2_W  s^2_w }{3\,v^2\,c^2_w} f(x_u,x_Z)
+
\frac { 4 \pi\alpha }{3 }  f(x_s,x_\tau)\nonumber\\
&-& \frac { 2 m^2_W  s^2_w }{\,v^2\, c^2_w}
(-\frac{1}{2}+\frac{1}{3}s^2_w) h(x_s,x_\tau,x_Z)\bigg)
\bigg(\frac{m_u V^\star_{us} }{v_u} - \sum\limits_{j = 1}^3
V^\star_{j2}
\epsilon^{ u}_{j1}\tan\beta \bigg)\nonumber\\
&+& \frac{ v\, m_\tau \sin\beta\,\tan\beta}{ 4\pi^2 m^2_H
V^\star_{us} }\bigg(16 \pi^2 + \frac { 8\,\pi \alpha }{3 }
f(x_u,x_\tau)-\frac {  m^2_W }{\,v^2\,c^2_w}
(\frac{1}{2}-\frac{2}{3}s^2_w) f(x_u,x_Z) + \frac { 4 \pi\alpha
}{3 }  f(x_s,x_\tau)\nonumber\\&-& \frac { 2 m^2_W
 s^4_w }{3\,v^2\, c^2_w} h(x_s,x_\tau,x_Z) \bigg)\bigg( \frac{m_s
V^\star_{us}}{v_d} - \sum\limits_{j = 1}^3 V^\star_{1j} \epsilon^{
d\star}_{j2}\tan\beta \bigg)+\triangle_Z+\triangle_\gamma,\nonumber\\
C^{H^\pm}_T &=& \frac{ v\, m_\tau \sin\beta}{ 4\pi^2 m^2_H
V^\star_{us}}\bigg(\frac { 2 m^2_W  s^2_w  }{3 v^2 c^2_w}
f(x_u,x_Z)- \frac { m^2_W  (-\frac{1}{2}+\frac{1}{3}s^2_w) }{ v^2
c^2_w}
f(x_s,x_Z)  + \frac{ 4 \pi\alpha }{3 }  f(x_s,x_\tau)\nonumber\\
&-& \frac { 2 m^2_W \, s^2_w }{\,v^2\, c^2_w}
(-\frac{1}{2}+\frac{1}{3}s^2_w) h(x_s,x_\tau,x_Z)+\frac { 8\pi
\alpha }{3} f(x_u,x_\tau)\bigg) \bigg(\frac{m_u V^\star_{us}
}{v_u} - \sum\limits_{j = 1}^3 V^\star_{j2} \epsilon^{
u}_{j1}\tan\beta \bigg)\nonumber\\ &-& \frac{ v\, m_\tau
\sin\beta\,\tan\beta}{ 4\pi^2 m^2_H V^\star_{us} }\bigg(\frac {
m^2_W  }{\,v^2\,c^2_w} (\frac{1}{2}-\frac{2}{3}s^2_w) f(x_u,x_Z)+
\frac {  m^2_W  s^2_w}{3\,v^2\,c^2_w} f(x_s,x_Z) - \frac{ 4
\pi\alpha }{3 }  f(x_s,x_\tau)\nonumber\\
&+&\frac { 2 m^2_W  s^4_w }{3\,v^2\, c^2_w} h(x_s,x_\tau,x_Z)-
\frac { 8\,\pi \alpha }{3} f(x_u,x_\tau)\bigg) \bigg( \frac{m_s
V^\star_{us}}{v_d} - \sum\limits_{j = 1}^3 V^\star_{1j} \epsilon^{
d\star}_{j2}\tan\beta \bigg), \label{Higgsw}
\end{eqnarray}
where $v$ denotes the Higgs vacuum expectation value which is
defined as $v=\sqrt{v^2_u+v^2_d}=\sqrt{\frac{2 m^2_W}{g^2}}\simeq
174\, GeV$. The terms $\triangle_Z$ and $\triangle_\gamma$
represent the contributions to $C^{H^\pm}_S $ originating from the
exchange of $V=Z,\gamma$ boson between allowed two lines, $A$ and
$B$, at each vertex of the tree-level diagram mediated by $H^\pm$
exchange. Together, with the box contributions, the gauge
independence results is guaranteed at one loop-level considered
here. The explicit expressions of $\triangle_Z$ and
$\triangle_\gamma$ are listed in the Appendix. The integration
loop functions are given, in terms of $ x_i =\frac{m_i^2}{m^2_H}
$, as \bea g(x_i,x_j,x_k) &=& \frac{x_i log
x_i}{(x_i-1)(x_i-x_j)(x_i-x_k)}+\frac{x_j log
x_j}{(x_j-1)(x_j-x_i)(x_j-x_k)}\nonumber\\&+&\frac{x_k log
x_k}{(x_k-1)(x_k-x_i)(x_k-x_j)},\nonumber\\
k(x_i,x_j)&=& \frac{-1}{(x_i-x_j)}\Big(\frac{1}{1-x_i}log
x_i-(x_i\leftrightarrow x_j)\Big),\nonumber\\
f(x_i,x_j)&=& \frac{1}{(x_i-x_j)}\Big(\frac{x_i}{1-x_i}log
x_i-(x_i\leftrightarrow x_j)\Big),\nonumber\\
h(x_i,x_j,x_k) &=&\frac{x^2_i log
x_i}{(x_i-1)(x_i-x_j)(x_i-x_k)}+\frac{x^2_j log
x_j}{(x_j-1)(x_j-x_i)(x_j-x_k)}\nonumber\\&+&\frac{x^2_k log
x_k}{(x_k-1)(x_k-x_i)(x_k-x_j)},\eea

It should be noted that an investigation of the dependency of the
matrix elements of $H\to \gamma\gamma$ through one $W^\pm$ loop in
SM on the choice of the gauge was done in Ref.\cite{Wu:2017rxt}.
The results showed that the calculated matrix elements using
$R_\zeta$ gauge and the unitary gauge are explicitly verified to
be different. However, in a subsequent study, it was found that
the unitary gauge is consistent and equivalent to the $R_\zeta$
gauge at the level of $\beta$-functions \cite{Irges:2017ztc}. At
higher loops, their is a possibility that the observed consistency
between the two gauges may breaks down. In this case, their is a
demand of a better understanding of the quantum internal structure
of spontaneously broken gauge theories \cite{Irges:2017ztc}. In
Ref. \cite{Melnikov:2016nvo}, the mismatch which was pointed out
by S. L. Wu and T. T. Wu group is resolved by a dedicated
calculation of the unitary gauge. In a recent study in
Ref.\cite{Gallagher:2020ajd}, it was shown that the unitary gauge
is the default gauge for massive gauge bosons.

In order to estimate the contributions of the charged Higgs to the
amplitude of the decay process under consideration we need to
discuss the constraints imposed on the couplings
$\epsilon^{u,d}_{ij}$ appear in the expressions of $C^{H^\pm}_i$
above. The couplings $\epsilon^{d}_{12}$, and $\epsilon^{d}_{32}$
are stringently constrained from flavor changing neutral current
processes, in the down quark sector, due to the tree-level neutral
Higgs exchange \cite{Crivellin:2012ye,Crivellin:2013wna}. On the
other hand, the coupling $\epsilon^{d}_{22}$ can be strongly
constrained upon applying the naturalness criterion of 't Hooft to
the quark masses that reads \cite{Crivellin:2013wna}

\begin{eqnarray}
|v_{u(d)} \epsilon^{d(u)}_{ij}|&\leq&
\left|V^{CKM}_{ij}\right|\,{\rm max
}\left[m_{d_i(u_i)},m_{d_j(u_j)}\right]\,\,\,\,\,\,\,\, for\, i <
j,\nonumber\\
|v_{u(d)} \epsilon^{d(u)}_{ij}|&\leq&{\rm max
}\left[m_{d_i(u_i)},m_{d_j(u_j)}\right]\, \,\,\,\,\,\,\,\,
\,\,\,\,\, \,\,\,\,\, \,\,\,\,\, \,\,\,\,\,  for\, i \geq
j,\label{constr}
\end{eqnarray}

Clearly, from this bound, $\epsilon^d_{22}$ is severely
constrained by the smallness of the $s$ quark mass. As a result,
we can safely neglect the contributions of the couplings
$\epsilon^{d}_{ij}$ to  $C^{H^\pm}_1,C^{H^\pm}_4$. Other terms, in
these Wilson coefficients are real and thus are not relevant for
generating $CP$ asymmetries. Thus to a good approximation we can
write
\begin{eqnarray}
C^{H^\pm}_{V}&\simeq& \frac { 8 m^2_W m_s m^2_\tau \sin^2\beta
s^2_w }{3 v\, V^\star_{us}\,c^2_w \cos\beta } \bigg((-\frac{1}{2}+
s^2_w)\, g(x_s,x_\tau,x_Z) + \frac { 4 v^2\,\pi \alpha \, m_u
c^2_w }{ m^2_W m_s s^2_w } k(x_u,x_\tau)
\bigg)\big(V^*_{cs}\, \epsilon^{ u}_{21} + V^*_{ts}\, \epsilon^{ u}_{31}\big),\nonumber\\
C^{H^\pm}_S &\simeq& -\frac{4 v\, m_\tau \sin^2\beta
}{V^\star_{us}\,m^2_H \cos\beta }\big(V^*_{cs}\,
\epsilon^{ u}_{21} + V^*_{ts}\, \epsilon^{ u}_{31}\big), \nonumber\\
C^{H^\pm}_T &\simeq& -\frac{4 v m_\tau \sin^2\beta }{V^\star_{us}
m^2_H \cos\beta}\bigg(\frac { 2 m^2_W m^2_H s^2_w }{3 v^2 c^2_w}
f(x_u,x_Z)- \frac { 2 m^2_W m^2_H s^2_w }{v^2\, c^2_w}
(-\frac{1}{2}+\frac{1}{3}s^2_w)
h(x_s,x_\tau,x_Z)+\frac { 8\pi \alpha \,m^2_H}{3} f(x_u,x_\tau)\nonumber\\
&+& \frac{ 4 \pi\alpha m^2_H}{3 } f(x_s,x_\tau) - \frac { m^2_W
m^2_H (-\frac{1}{2}+\frac{1}{3}s^2_w) }{ v^2 c^2_w} f(x_s,x_Z)
\bigg)\big(V^*_{cs}\, \epsilon^{ u}_{21} + V^*_{ts} \epsilon^{
u}_{31}\big).\label{Higgsw}
\end{eqnarray}

It should be noted that, in the above expressions, we neglected
non relevant real terms. In addition we neglected terms
proportional to $\epsilon^{u}_{11}$ which is severely constrained
from the bound in Eq.(\ref{constr}) due to the smallness of the up
quark mass.

 Recently, a lower bound $ m_{H^\pm}\gtrsim 600$ GeV, independent
of $\tan \beta$, has been obtained in 2HDM II  after taking into
account all relevant results from direct charged and neutral Higgs
boson searches at LEP and the LHC,  as well as the most recent
constraints from flavour physics \cite{Arbey:2017gmh}. This bound
should be also respected in 2HDM III \cite{Crivellin:2012ye}.
Thus, for $ m_{H^\pm} = 600$ GeV and $\tan \beta = 50$ we find
that

\begin{eqnarray}
C^{H^\pm}_{V}&\simeq& - 5.23\times 10^{-6} \big(V^*_{cs}\,
\epsilon^{u}_{21} + V^*_{ts}\, \epsilon^{ u}_{31}\big),\nonumber\\
C^{H^\pm}_S &\simeq& - 0.76 \big(V^*_{cs}\, \epsilon^{
u}_{21} + V^*_{ts}\, \epsilon^{ u}_{31}\big),\nonumber\\
C^{H^\pm}_T &\simeq& 7.49 \,\times 10^{-3} \big(V^*_{cs}\,
\epsilon^{ u}_{21} + V^*_{ts}\, \epsilon^{ u}_{31}\big),
\label{Higgsw}
\end{eqnarray}

Clearly, to a good approximation, we can neglect contribution of
$C^{H^\pm}_{V}$ to the amplitude as it is very small. The previous
equation shows that $C^{H^\pm}_S $ is much larger than
$C^{H^\pm}_T$. However, as we discussed before, only $C^{H^\pm}_T$
can generate non vanishing direct CP asymmetry.

The 't Hooft naturalness criterion leading to the bounds  in
Eq.(\ref{constr}) implies different bounds for $\epsilon^u_{ij}$
and $\epsilon^u_{ji}$ due to the presence of the CKM element in
the first line of Eq.(\ref{constr}) corresponding to the case $i <
j$. Thus one expects different bounds for $\epsilon^u_{12}$ and
$\epsilon^u_{21}$ as can be seen from Eq.(16) in
Ref\cite{Crivellin:2013wna}. The authors of
Ref.\cite{Crivellin:2013wna} set a bound on $\epsilon^u_{21}$
after considering all constraints listed in their Tables VII and
VIII. This bound can be read from the first matrix in Eq.(75) as
$\mid\epsilon^u_{21}\mid\leq 3.0\times 10^{-2}$. However, the
strong bound $\mid\epsilon^u_{12}\epsilon^{u\,*}_{21}\mid\ <
2\times 10^{-8}$ from considering $D-\bar D$ mixing
\cite{Crivellin:2013wna} implies that $\mid\epsilon^u_{12,21}\mid
< \sqrt{2}\times 10^{-4}$. This can be understood as  in the
absence of a symmetry that protect $\mid\epsilon^u_{12}\mid$ from
being much smaller than $\mid\epsilon^u_{21}\mid$ and using the
same merit of the 't Hooft naturalness criterion it is  unnatural
to have $\mid\epsilon^u_{12}\mid$ much suppressed than
$\mid\epsilon^u_{21}\mid$. Thus we obtain
\begin{eqnarray}
C^{H^\pm}_S &\simeq& - 0.76  V^*_{ts}\, \epsilon^{ u}_{31}, \nonumber\\
C^{H^\pm}_T &\simeq& 7.49 \,\times 10^{-3}  V^*_{ts}\, \epsilon^{
u}_{31}. \label{Higgsw}
\end{eqnarray}
Knowing that $Re(V_{ts})\simeq{\mathcal O} (10^{-2})$ and the
bound $ 2.7\times 10^{-3} \leq \mid\epsilon^u_{31}\mid \leq
2.0\times 10^{-2}$ from the process $B\to \tau \nu$
\cite{Crivellin:2013wna}, it is clearly trivial that the  $B\to
\tau \nu$  process sets more severe bound on $\epsilon^u_{31}$
compared to the model-independent EDM bound. Consequently, one
expects a strong bound $Im \, C^{H^\pm}_T < 10^{-6}$ and after
using CKM matrix elements from Ref \cite{utfit}, we obtain from
Eq.(\ref{ACP}) \be \big|A^{H^\pm}_{CP}\big|\lesssim {\mathcal O}
(10^{-8}), \label{cT}\ee Clearly, the charged Higgs contributions
can enhance the direct CP asymmetry several orders of magnitude
larger than the standard model prediction. However, the estimated
CP asymmetry is still so small to be detected by current or near
future experiments.

\subsection{ $\tau^-\to K^-\pi^0\nu_\tau $  in Leptoquarks
models.}\label{Leptoquark}

 Leptoquark particles are scalars or vectors bosons that have both baryon  and lepton number \cite{Pati:1974yy,Buchmuller:1986zs}.
 They appear for instance in grand unified theories (GUTs) \cite{Langacker:1980js,Frampton:1991ay,Hewett:1988xc},
 technicolor models \cite{Farhi:1980xs,Lane:1991qh}  and SUSY models with R-parity violation. At low energy,
 Leptoquarks can be  described as an effective four fermion interaction induced by leptoquark exchange.
 Several observables have been used to set bounds on these effective couplings
as is the case of D meson decays
\cite{Davidson:1993qk,Dobrescu:2008er,Dorsner:2009cu} and K
mesons.

Scalar leptoquarks $S$ may couple to both left or right handed
quark chiralities.  Let us consider the exchange of the following
scalar leptoquarks:
\begin{itemize}
\item  $S_{1/2}$ with charge $2/3$ and SU(3)$_{\rm
C}\times$SU(2)$_{\rm L}\times$U(1)$_{\rm Y}$ gauge numbers given
as $(3,2,7/3)$; and \item the $S_{0}$ with charge $-1/3$ and
$(3,1,-2/3)$ gauge numbers.
\end{itemize}

Their Yukawa couplings with SM fermions are given by
\begin{eqnarray}
&&{\cal L}_{S_{1/2}}=[\kappa^L_{ij}\bar{U}_{R_i}L_{L_j}
-\kappa^R_{ij}\bar{Q}_{L_i}i\tau_2 E_{R_j}] S_{1/2}+{\rm h.c.},\nonumber\\
&&{\cal L}_{S_{0}}=[\xi^L_{ij}\bar{Q}_{L_i}i\tau_2L^c_{L_j}
+\xi^R_{ij}\bar{U}_{R_i}E^c_{R_j}] S_{0}+{\rm h.c.},\label{YKC}
\end{eqnarray}

Here, $\tau_i$ ($i=1,2,3$) are the Pauli matrices,
$\bar{Q}_{L_i}=(\bar{u}_i,\bar{d}_i)_L$ and
$L_{L_i}=(\bar{\nu}_i,\bar{e}_i)_L$  and  for a spinor field
$\psi$, the charge conjugation $\psi^c$ is defined as
$\psi^c_{R,L}=i\gamma^0\gamma^2\bar{\psi}^T_{R,L}$. The coupling
constants $\kappa^{L,R}_{ij}$ and $\xi^{L,R}_{ij}$ can be in
general complex numbers. The Yukawa couplings, in Eq.(\ref{YKC}),
lead to the Lagrangians
\begin{eqnarray}
&&{\cal L}^{I}=\left[\kappa^L_{1i}
\bar{u}_{R}\nu_{i_L}+\kappa^R_{23}\bar{s}_{L}\tau_{R}\right]
S_{1/2}
+{\rm h.c.},\nonumber\\
&&{\cal L}^{II}=\left[-\xi^L_{13}\bar{u}_{L}\tau^c_{L}
+\xi^L_{2i}\bar{s}_{L}\nu^c_{i_L} +
\xi^R_{13}\bar{u}_R\tau^c_{R}\right]S_0 + {\rm
h.c.},\label{LeptoquarkLagrangian}
\end{eqnarray}
which is relevant to $\tau\rightarrow K^-\pi^0\nu$ decay with
$i=1,2,3$. These Lagrangians result in diagrams similar to the one
mediated by the charged Higgs in which the Higgs boson is replaced
with the scalar leptoquarks $S_{1/2}$ and $S_{0}$. After
integrating out the leptoquarks, we can obtain the Wilson
coefficients $C^{I,II}_i$, corresponding to the operators $Q_i$ in
Eq.(\ref{Qi}), as
\begin{eqnarray}
 C_{V}^{I}&=&0,\,\,\,\,\,\,\,\,\,\,\,\,\,\,\,\,\,\,
 C_{S}^{I}=\frac{\kappa^R_{23}
\kappa^{L*}_{1i}}{4 \sqrt{2}\, G_F V^*
_{us}\,M^2_{S_{1/2}}},\,\,\,\,\,\,\,\,\,\,\,\,\,\,\,\,\,\,
C_{T}^{I}=\frac{1}{4} C_{3}^{I},\nonumber\\
 C_{V}^{II}&=& \frac{\xi^L_{2i} \xi^{L *}_{13}}{4 \sqrt{2}\, G_F V^* _{us}
 M^2_{S_{0}}},\,\,\,\,\,\,\,\,\,\,\,\,\,\,\,\,\,\,
 C_{S}^{II}=\frac{ \xi^L_{2i} \xi^{R *}_{13}}{4 \sqrt{2}\, G_F V^* _{us}
M^2_{S_{0}}}\,\,\,\,\,\,\,\,\,\,\,\,\,\,\,\,\,\,
C_{T}^{II}=\frac{1}{4} C_{S}^{II}.\label{leptq}
\end{eqnarray}
We now discuss the constraints imposed on the leptoquark
couplings. For model $II$, the couplings $|\xi^L_{2i} \xi^{L
*}_{13}|$, appears in $ C_{V}^{II}$, can be severely constrained
from the process $K\to \pi\bar\nu\nu$ \cite{Davidson:1993qk}. From
the results of the analysis carried in Ref.\cite{Davidson:1993qk},
the bound reads $|\xi^L_{2i} \xi^{L *}_{13}|< 2\times
10^{-5}\times (M_{S_{0}}/[100\,GeV])^2$. Thus, in  model $II$, we
have $C_V= C^{SM}_V + C_{V}^{II}\simeq C^{SM}_V$.

Turning now to the couplings $ \xi^L_{2i} \xi^{R *}_{13}$ and
setting $i=3$, we find that possible constraints can be derived
from the observable $\tau\to K\nu_\tau$. The expected bound on $
\xi^L_{23} \xi^{R *}_{13}\equiv \xi^L_{s \nu_\tau} \xi^{R *}_{u
\tau}$, which corresponds to the Wilson coefficients of the four
fermion operator that contributes to both $\tau\to K\nu_\tau$ and
$\tau^-\to K^-\pi^0\nu_\tau $, read $Re(\xi^L_{23} \xi^{R
*}_{13})\sim [-0.07,0.04]$ and $|Im(\xi^L_{23} \xi^{R
*}_{13})|\sim [0.0,0.7]$ for $M_{S_{0}}=1 \,TeV$
\cite{deBoer:2017que}. Thus, for $i=3$ in model $II$, we have
$|Im(C_T)|=|Im( C_{T}^{II})|\sim [0.0,0.01]$. It should be noted
that a strong constraint on $|Im( C_{T}^{II})|$ can be obtained
from the EDM of the neutron noticing that the coupling $\xi^L_{23}
\xi^{R *}_{13}$ can be linked to $C_{3321}\equiv C_{\nu_\tau \tau
s u}$, defined in Eq.(\ref{eq:LT1}), and using the bound given in
Eq.(\ref{ctc}). As a consequence, \be
\big|A^{II}_{CP}\big|\lesssim {\mathcal O} (10^{-7}),
\label{cT}\ee  similarly to the obtained value in the charged
Higgs model.

Regarding model $I$, we find that the predicted CP asymmetry will
be the same as in model $II$. This can be explained as similar
constraints can be imposed on the couplings $\kappa^R_{23}
\kappa^{L*}_{13}$ from the observable $\tau^-\to K^-\nu_\tau$
\cite{deBoer:2017que} and from the electric dipole moment (EDM) of
the neutron in Eq.(\ref{ctc}).

In the scalar leptoquark models discussed above, the amplitudes of
the processes $\tau^-\to K^-\pi^0\nu_{e}$ and $\tau^-\to
K^-\pi^0\nu_{\mu}$ receive contributions from the couplings $
\xi^L_{21} \xi^{R *}_{13} \equiv \xi^L_{s \nu_e} \xi^{R *}_{u
\tau} $ and $ \xi^L_{22} \xi^{R *}_{13} \equiv \xi^L_{s \nu_\mu}
\xi^{R *}_{u \tau}$ respectively. These couplings can be also
constrained using the observable $\tau\to K\nu$
\cite{Carpentier:2010ue}. However, the obtained bounds turns to be
one order of magnitude weaker than the bound on $ \xi^L_{23}
\xi^{R *}_{13}\equiv \xi^L_{s \nu_\tau} \xi^{R *}_{u \tau}$.
Moreover, the couplings $ \xi^L_{21} \xi^{R *}_{13} \equiv
\xi^L_{s \nu_e} \xi^{R *}_{u \tau} $ and $ \xi^L_{22} \xi^{R
*}_{13} \equiv \xi^L_{s \nu_\mu} \xi^{R *}_{u \tau}$ cannot be
linked to $C_{3321}\equiv C_{\nu_\tau \tau s u}$, defined in
Eq.(\ref{eq:LT1}), and thus evade the strong bound given in
Eq.(\ref{ctc}). Consequently, we expect a somehow large branching
ratios and CP asymmetries for the channels $\tau^-\to
K^-\pi^0\nu_e$ and $\tau^-\to K^-\pi^0\nu_\mu $ compared to their
corresponding ones of $\tau^-\to K^-\pi^0\nu_{\tau}$. Since these
couplings contribute to $C_{S,T}$ one expects that the effect on
the branching ratios is not much while a large effect is expected
to be seen in the CP asymmetries. With the only constraints from
 $\tau\to K\nu$ and in the absence of much stronger constraints, the predicted
direct CP asymmetry of $\tau^-\to K^-\pi^0\nu_e$ and $\tau^-\to
K^-\pi^0\nu_\mu $  can be of order $10^{-3}$.

 Finally, we discuss the possibility of having constraint on the $Im\, C_T$, in the leptoquark models
discussed in this work, from the process $B \to \tau \nu$. Recall
that, the process $B\to\tau\nu$ can be generated through $b\to u
\tau \nu$ transition. In model $II$, the transition originates at
tree-level from the exchanging of the scalar leptoquark $S_0$. The
relevant product of the leptoquark couplings to the amplitude of
$B\to\tau\nu$ decay is  $ \xi^{L*}_{3i}\, \xi^{R}_{13}$.
Consequently, the  constraints from $B\to\tau\nu$ process are
expected to be imposed on $ \xi^{L*}_{3i}\, \xi^{R}_{13}$ which is
irrelevant to $\tau\rightarrow K^-\pi^0\nu$ process that receives
contributions from $ \xi^L_{2i} \xi^{R *}_{13}$ as we have shown
above. Regarding  model $I$, first line in
Eq.(\ref{LeptoquarkLagrangian}), the $b\to u \tau \nu$ transition
can be generated at tree-level from the exchanging of the scalar
leptoquark $S_{1/2}$. In this case, the relevant leptoquark
couplings to the amplitude of $B\to\tau\nu$ decay are $
\kappa^{R*}_{33} \kappa^L_{1i} $. Again, the expected constraints
from $B\to\tau\nu$ process  on $ \kappa^{R*}_{33} \kappa^L_{1i} $
are irrelevant to $\tau\rightarrow K^-\pi^0\nu$ process that
receives contributions from $\kappa^R_{23} \kappa^{L*}_{1i}$.

\section{Nonintegrated  CP violation asymmetries  \label{lcpv}}

Searches for local CPV signal have been done by Cleo
\cite{Bonvicini:2001xz} and Belle \cite{Bischofberger:2011pw}
collaborations to obtain bounds on new generic scalar mediators
like the ones considered here. Belle collaboration is one order of
magnitude more precise. They parameterized new physics with
$\eta_S\simeq C_S$ by doing the replacement $f_0\to [1-\eta_S
s/(m_\tau(m_s-m_u))]f_0$. No local CPV signal was found and a
bound of $|{\rm Im}\ (\eta_S)|\simeq |{\rm Im}\ (C_S)|<0.013$ was
obtained. Recall that within SM, at tree-level, $C^{SM}_S=0$ which
is not the case in some beyond SM physics that have been
investigated in the previous section.  In the case of 2HDM III,
and using Eq.(\ref{Higgsw}), this bound results in  the
constraints $|{\rm Im}\ (\epsilon^{ u}_{31})|<0.4$. On the other
hand and in the case of scalar Leptoquark,  the  bound  $|{\rm
Im}\ (\eta_S)|\simeq |{\rm Im}\ (C_S)|<0.013$ will lead to a
weaker bound on $|{\rm Im}\ (C^{I,II}_S)|$ compared to that one
obtained from the electric dipole moment (EDM) of the neutron in
Eq.(\ref{ctc}) as, from Eqs.(\ref{ctc},\ref{leptq}), we have
$|{\rm Im}\ (C^{I,II}_S)|\lesssim 4\times 10^{-5}$.

\begin{figure}[tbhp]
\includegraphics[width=7cm,height=6cm]{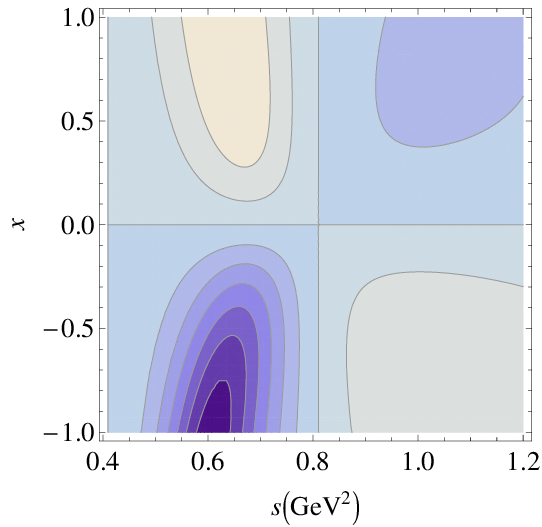}
\hspace{0.05cm}
\includegraphics*[width=1cm,height=6cm]{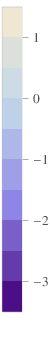}
\hspace{0.1 cm}
\includegraphics*[width=7cm,height=6cm]{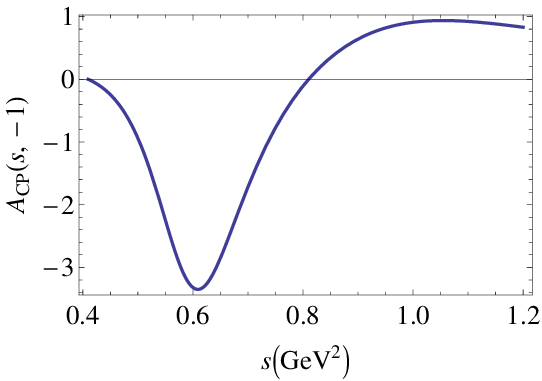}
\medskip
\caption{Local CPV, in units of $10^{-3}$ as a function of $s$ and
$x=\cos \theta$ and the same but for $x=-1$. \label{localcpv}}
\end{figure}

We proceed now to calculate the local CPV given in Eq.(\ref{LCPV})
within 2HDM III discussed above. Including charged Higgs
contributions to the Wilson coefficients $C_V$ and $C_S$ in
Eq.(\ref{LCPV}) can be done via the replacement $C_V\to
C^{SM}_1\simeq 1$, $C_S \to 1+C^{H^\pm}_S s/(m_\tau m_s)\simeq 1-
0.76  V^*_{ts}\, \epsilon^{ u}_{31} s/(m_\tau m_s)$. For a value
of $Im(\epsilon^{ u}_{31})\leq 2 \times 10^{-2}$ that satisfies
the strongest bound from $B \to \tau \nu$ process we find that
$Im(C_S)\lesssim 3.6\times 10^{-3}\times s$.

We turn now to the form factors $f_+(s)$ and $f_0(s)$  required to
the calculation of the local CPV.  Recall that a discussion about
the form factor $f_+(s)$ has been presented in the preceding
section. So we are left here to discuss the scalar form factor
$f_0$.  While the phase of $f_0$ was measured by the LASS
collaboration \cite{pelaez} its magnitude can be obtained by using
dispersion relations \cite{Moussallam:2007qc} but uncertainties in
the input information avoid a precise prediction. In fact decays
as the one we are interested in here should provide experimental
information about $f_0$, once the angular distribution can be
measured \cite{Kimura:2014wsa}. For our purpose of an estimation
of the local CPV the form factor $f_0$  can be parameterized, in
terms of a superposition of Breit-Wigner resonances following
Refs. \cite{Epifanov:2007rf,Dhargyal:2016kwp} as
\begin{eqnarray}
f_0(s)  &=& \chi {s\over m_{K_0^*(700)}^2 } BW(K^*(700)) +  \gamma
{s\over m_{K_0^*(1430)}^2 } BW(K^*(1430))
\nonumber \\
BW(R) &=& {m_R^2 \over m_R^2-s+i \sqrt{s}\ \Gamma_R(s) },\
\Gamma_R(s)= \Gamma_{0R} {m_R^2\over s} \left( {p(s)\over
p(m_R^2)} \right)^{2J+1}   \label{ffa}
\end{eqnarray}
with $p=|{\bf p}_K|$, $\chi=2.28$ and $\gamma=1.92{\rm
e}^{4.03\cdot i}$. Breit-Wigner parameters are
$m_{K_0^*(700)}=878(23)^{+64}_{-51}$,
$\Gamma_{K_0^*(700)}=499(52)^{+55}_{-87}$, $m_{K_0^*(1430)}=1425$
and $\Gamma_{K_0^*(1430)}=270$. The spin of the resonance is $J$
and  all masses and widths are expressed in MeV \cite{pdg}. With
all this in hand, we present a typical graphical representation of
the local CPV  in Figure \ref{localcpv}. Clearly, from Figure
\ref{localcpv}, values as large as 0.3 \%  can be obtained. Due to
the strong constraints

\section{Conclusion \label{sec:conclusion}}

In this paper we have derived the contributions to the effective
Hamiltonian governing the semileptonic $\vert\Delta S \vert=1$ tau
decays in 2HDM III and models with scalar Leptoquarks. We have
discussed the imposed constraints on the elements in the parameter
space of the model relevant to the decay channel $\tau^-\to
K^-\pi^0\nu_\tau$. In addition, we have analyzed the role of the
different contributions, originating from the scalar, vector and
tensor hadronic currents, in generating direct CP asymmetry in the
decay rate of $\tau^-\to K^-\pi^0\nu_\tau$.

 We have shown that non vanishing direct CP asymmetry in the decay
rate of $\tau^-\to K^-\pi^0\nu_\tau$ can be generated in the model
due to the presence of the weak phase in the Wilson coefficient
$C^{NP}_T$ and due to the strong phase difference resulting from
the interference between the form factors $B_T(s)$, and $f_+(s)$.
After taking into account the relevant constraints we found that
the resultant CP asymmetry can be of order $10^{-8}$ in both
models. The asymmetry is so tiny to be probed even in near future
experiments.

In this work we have also studied another observable related to
the CP violation, namely the local CPV. This kind of asymmetry can
be generated if there is interference between vector and scalar
contributions, as long as they contribute with different weak
phases which is not the case in the SM. In the 2HDM III we studied
in this work, we have found  that, direct local CPV can be as
large as $0.3$ \% not far from experimental possibilities.

\section*{Acknowledgements}
The D.D.'s work was partially support by CONACYT project CB-
286651 and Conacyt-SNI, and DAIP Project.

\appendix

\section{\label{app}}

The quantities $\triangle_Z$ and $\triangle_\gamma$  can be
expressed as

 \bea \triangle_Z &=& a^{u s }_Z+a^{u H }_Z+ a^{s H }_Z
+a^{\tau\nu_\tau}_Z +a^{\tau H}_Z +a^{\nu_\tau H}_Z,\nonumber\\
\triangle_\gamma &=& a^{u s }_\gamma+a^{u H }_\gamma+ a^{s H
}_\gamma+a^{\tau H}_\gamma, \eea

Here, $a^{AB}_V$ denotes the contribution from Feynman diagram
where the gauge boson $V=Z,\gamma$ connect the two lines $A$ and
$B$ at one vertex of the tree-level diagram mediated by $H^\pm$
exchange. The results of each  $a^{AB}_V$ can be written in terms
of Passarino -Veltman (PV) functions where the sum of the
divergent parts, of some of these functions, in the total
amplitude vanishes. In the following we show only dominant terms
where, in the coefficients of each PV functions, terms
proportional to $m_u$, $m_s$, $s$ and $m_\tau$ can be neglected
compared to the terms proportional to $m_Z$ and $m_H$. Defining
$g^H_L=\big(\Gamma_{u s }^{H^\pm\,LR\,\rm{eff} } \big)^*$,
$g^H_R=\big(\Gamma_{u s }^{H^ \pm\,RL\,\rm{eff} } \big)^*$,
$g^\tau_L=-\frac{1}{2}+s^2_w$, $g^\tau_R=s^2_w$,
$g^u_L=\frac{1}{2}-\frac{2}{3} s^2_w$, $g^u_R=-\frac{2}{3}s^2_w$,
$g^s_L=-\frac{1}{2}+\frac{1}{3}s^2_w$ and $g^s_R=\frac{1}{3}s^2_w$
we find that the different contributions to $\triangle_Z$ can be
expressed as

\begin{eqnarray} a^{u s  }_Z &\simeq& - \frac{m_\tau\,m^2_Z\tan\beta}
{16 \pi ^2 \, s \,v\, V^\star_{us}\left(s-m_{H^\pm
}^2\right)}\bigg( 4 m_s^2\bigg[ g_L{}^H g_L{}^s g_L{}^u+g_R{}^H
g_R{}^s g_R{}^u\bigg]
B_0\left(m_s^2+m_u^2-s,m_s^2,m_u^2\right)\nonumber\\
&+& \bigg[g_L{}^H g_L{}^u \left(3\, s \, g_R{}^s-4 m_s^2
g_L{}^s\right)+g_R{}^H g_R{}^u \left(3 \,s\, g_L{}^s-4 m_s^2\,
g_R{}^s\right)\bigg]B_0\left(m_s^2,m_s^2,m_Z^2\right)\nonumber\\
&+&4s\big[g_L{}^H g_L{}^u g_R{}^s+g_R{}^H g_L{}^s g_R{}^u\big]
B_0\left(0,m_u^2,m_Z^2\right) +  s \big(g_R{}^H g_R{}^u-g_L{}^H
g_L{}^u\big)\big(g_L{}^s-g_R{}^s\big)
B_0\left(0,m_s^2,m_Z^2\right)  \nonumber\\&-& \frac{2 s}{m_Z^2}
\big(g_L{}^H g_L{}^u g_R{}^s+g_R{}^H g_L{}^s g_R{}^u\big) A_0
(m_Z^2) +\bigg[4 m_s^2 m_Z^2\big( g_L{}^H g_L{}^s g_L{}^u+g_R{}^H
g_R{}^s g_R{}^u\big)\nonumber\\&+& 4 s^2 \big(g_L{}^H g_L{}^u
g_R{}^s+g_R{}^H g_L{}^s g_R{}^u\big)\bigg]
C_0\left(m_s^2,m_u^2,m_s^2+m_u^2-s,m_s^2,m_Z^2,m_u^2\right)\bigg),
\nonumber\\ a^{u H }_Z &\simeq& -\frac{  m_\tau\,m_Z\tan\beta (1-2
s^2_w) (g_L{}^H  g_L{}^u+g_R{}^H  g_R{}^u)}{16 \pi ^2 V^\star_{us}
\left(s-m_{H^\pm }^2\right)}\bigg(-4 B_0\left(m_s^2+4 m_u^2+2
s,m_H^2,m_u^2\right) -\frac{1}{m_Z^2}
\text{A}_0\left(m_Z^2\right)\nonumber\\&+& \frac{3}{2}
B_0\left(m_u^2,m_u^2,m_Z^2\right)+\left(2 m_H^2 -4
m_Z^2\right)\text{C}_0\left(m_u^2,m_s^2+m_u^2+s,m_s^2+4 m_u^2+2
s,m_u^2,m_Z^2,m_H^2\right)\nonumber\\&+& \left( 3 -
\frac{m_H^2}{m_Z^2} \right)
B_0\left(m_s^2+m_u^2+s,m_H^2,m_Z^2\right)+\frac{1}{2}
B_0\left(0,m_u^2,m_Z^2\right)\bigg), \nonumber\\
a^{s H }_Z &\simeq&-\frac{ m_\tau\,m_Z \tan\beta(1-2 s^2_w)
\left(g_L{}^H g_R{}^s +g_R{}^H g_L{}^s \right)}{16 \pi ^2
V^\star_{us}\left(s-m_{H^\pm }^2\right)}\bigg(-2 m_H^2
\text{C}_0\left(m_s^2,m_u^2,m_s^2+m_u^2+s,m_Z^2,m_s^2,m_H^2\right)\nonumber\\&+&\left(1+\frac{m_H^2}{m_Z^2
}\right) B_0\left(m^2_u+m^2_s+s,m_H^2,m_Z^2\right) -\frac{1}{2}
B_0\left(0,m_s^2,m_Z^2\right) -\frac{m_H^2 m_s^2}{s}
B_0\left(m_s^2,m_s^2,m_Z^2\right)
\nonumber\\&+&\frac{1}{m_Z^2}\text{A}_0\left(m_Z^2\right) \bigg),
\nonumber\\
 a^{\tau\nu_\tau}_Z &\simeq&\frac {
m_\tau\,m^2_Z\tan\beta \left(g_L{}^H  +g_R{}^H  \right) }{32 \pi
^2\,\,v\,V^\star_{us} \left(s-m_{\tau }^2\right)\left(s-m_{H^\pm
}^2\right)} \bigg( 4 \bigg[m_{\tau }^2 m_Z^2 g_L{}^{\tau
}+\left(m_{\tau }^2-s\right){}^2 g_R{}^{\tau
}\bigg]\text{C}_0\big(0,s,m_{\tau }^2,m_Z^2,0,m_{\tau
}^2\big)\nonumber\\&+& 4 m_{\tau }^2 g_L{}^{\tau }
B_0\left(s,0,m_{\tau }^2\right)+\frac{1}{2}\big(3 m_{\tau
}^2+s-8\, s\, g_R{}^{\tau }\big) B_0\big(m_{\tau }^2,m_Z^2,m_{\tau
}^2\big) +\frac{2\big(s-m_{\tau }^2\big) }{ m_Z^2}g_R{}^{\tau } \text{A}_0\left(m_Z^2\right)\nonumber\\
&-& 4 \big(s-m_{\tau }^2\big) g_R{}^{\tau }
B_0\left(0,0,m_Z^2\right)-\frac{1}{2}\big(s-m_{\tau
}^2\big)B_0\left(0,m_Z^2,m_{\tau }^2\right)\bigg),
\nonumber\\
a^{\tau H}_Z &\simeq&\frac{m_\tau\,m_Z \tan\beta (1-2
s^2_w) \left(g_L{}^H  +g_R{}^H  \right)}{32 \pi ^2\,V^\star_{us}
m_Z^2 \left(s-m_{H^\pm }^2\right)(s-m_{\tau }^2)} \bigg(
\bigg[m_{\tau }^2 m_H^2 -\frac{1}{2}m_Z^2 \left(m_{\tau }^2 + s
\right)\bigg]B_0\left(m_{\tau }^2,m_Z^2,m_{\tau
}^2\right)\nonumber\\&+& 2\bigg[m_{\tau }^2
\left(m_H^2-m_Z^2\right) g_L{}^{\tau }-s \left(m_H^2+m_Z^2\right)
g_R{}^{\tau }\bigg]B_0\left(s,m_H^2,m_Z^2\right) -\frac{1}{2}
\big(s-m_{\tau }^2\big)\nonumber\\
&\times& \bigg[  m^2_Z B_0\left(0,m_Z^2,m_{\tau }^2\right)-4
g_R{}^{\tau } \text{A}_0\left(m_Z^2\right)\bigg] + m_{\tau }^2
m_H^2(m_H^2-m_Z^2) \text{C}_0\left(0,s,m_{\tau }^2,m_{\tau
}^2,m_H^2,m_Z^2\right) \bigg),\nonumber\\
\end{eqnarray}

\begin{eqnarray} a^{\nu_\tau H}_Z &\simeq&
\frac{m_\tau\, m^2_Z \tan\beta (1-2 s^2_w)\left(g_L{}^H +g_R{}^H
\right)}{8\sqrt{2}\, \pi ^2\,v \,V^\star_{us} \left(s-m_{\tau
}^2\right)\left(s-m_{H^\pm }^2\right)} \bigg(2 \bigg[m_{\tau }^2
B_0\left(m_{\tau }^2,0,m_H^2\right)+\big(s\, m_H^2 -m_{\tau }^2
m_H^2+m_{\tau }^2 m_Z^2\big)\nonumber\\&\times&
\text{C}_0\left(0,s,m_{\tau }^2,0,m_Z^2,m_H^2\right)
\bigg]-\frac{1}{m_Z^2}\bigg[m_{\tau }^2 \left(m_Z^2-m_H^2\right)+s
\left(m_H^2+m_Z^2\right)\bigg]
B_0\left(s,m_H^2,m_Z^2\right)\nonumber\\&+&
\frac{1}{m_Z^2}\left(s-m_{\tau }^2\right)
\text{A}_0\left(m_Z^2\right)\bigg),
\end{eqnarray}
where we have used $g^\tau_L-g^\tau_R=-\frac{1}{2}$. Turning now
to the contributions to $\triangle_\gamma$ we find that

\begin{eqnarray}
 a^{s H}_\gamma &=&\frac{ \alpha _e m_\tau\,v\tan\beta \left(g_L{}^H  +g_R{}^H  \right)}{12 \pi\,V^\star_{us} \left(s-m_{H^\pm }^2\right)}
\bigg(B_0\left(s,0,m_H^2\right) -2 B_0\left(m_s^2,0,m_s^2\right)-
2 m_H^2 \text{C}_0
\left(m_s^2,m_u^2,s,0,m_s^2,m_H^2\right)\bigg),\nonumber\\
a^{u s}_\gamma &=&\frac{ \alpha _e m_\tau\,v\tan\beta
\left(g_L{}^H  +g_R{}^H  \right)}{9 \pi\,V^\star_{us}
\left(s-m_{H^\pm }^2\right)} \bigg(
s\text{C}_0\left(m_s^2,m_u^2,-s,m_s^2,0,m_u^2\right)+B_0\left(m_s^2,0,m_s^2\right)+B_0\left(m_u^2,0,m_u^2\right)
\bigg),\nonumber\\
a^{u H}_\gamma &=&-\frac{ \alpha _e m_\tau\,v\tan\beta
\left(g_L{}^H  +g_R{}^H \right)}{6 \pi
\,V^\star_{us}\left(s-m_{H^\pm }^2\right)} \bigg( 2
m_H^2\text{C}_0\left(m_u^2,s,2 s,m_u^2,0,m_H^2\right)+ 3
B_0\left(s,0,m_H^2\right)\nonumber\\&-& 4 B_0\left(2 s
,m_H^2,m_u^2\right)+2 B_0\left(m_u^2,0,m_u^2\right)
\bigg),\nonumber\\ a^{\tau H}_\gamma &\simeq&\frac{ \alpha _e
m_\tau\,v\tan\beta \left(g_L{}^H  +g_R{}^H  \right)}{4 \pi
\,V^\star_{us} \left(s-m_{\tau }^2\right)\left(s-m_{H^\pm
}^2\right)} \bigg(2\big(m_{\tau }^4- s \, m_H^2\big)
\text{C}_0\left(0,s,m_{\tau }^2,m_{\tau
}^2,m_H^2,0\right)+\left(m_{\tau }^2+s\right)
B_0\left(s,0,m_H^2\right)\nonumber\\&-& 2 s\, B_0\left(m_{\tau
}^2,0,m_{\tau }^2\right)\bigg).
\end{eqnarray}
The loop function $A_0(m^2)$ is defined as \be A_0(m^2)=
\frac{16\pi^2 \mu^{4-d}}{i}\int\frac {d^d k}{(2\pi
)^d}\frac{1}{k^2-\bar
m^2}=m^2\bigg[D+1-\ln\big(\frac{m^2}{\mu^2}\big)\bigg],\ee where
$\mu$ is the renormalization scale, $ \bar m^2=  m^2- i\epsilon$
and $D=2/(4-d)-\gamma_E+\ln (4\pi)$. Concerning the loop function
$B_0(\ell^2,m^2,n^2)$, it is defined as \bea B_0(p^2,m^2,n^2)&=&
\frac{16\pi^2 \mu^{4-d}}{i}\int\frac {d^d k}{(2\pi
)^d}\frac{1}{(k^2-\bar m^2)\big[(k+p)^2-\bar
n^2\big]}\bigg]\nonumber\\
&=& D-\int^1_0 dx\, \ln\bigg[\frac{-x(1-x) p^2+x n^2+(1-x)
m^2}{\mu^2}\bigg],\eea Defining $m^2_\ell=\frac{m^2}{\ell^2}$ and
$n^2_\ell=\frac{n^2}{\ell^2}$ we find that \bea
B_0(\ell^2,m^2,n^2)&=& D
-\ln\big(\frac{m^2}{\mu^2}\big)+2-\sqrt{-\lambda(1,m^2_\ell,n^2_\ell)}\bigg\{\sin^{-1}\bigg[\sqrt{{\frac{1}{4
m^2_\ell}}}\big(1+m^2_\ell-n^2_\ell\big)\bigg]\nonumber\\
&+& \sin^{-1}\bigg[\sqrt{{\frac{1}{4
n^2_\ell}}}\big(1+n^2_\ell-m^2_\ell\big)\bigg]\bigg\}+\frac{1}{2}\big(1+n^2_\ell-
m^2_\ell\big)\ln\big(\frac{n^2_\ell}{m^2_\ell}\big),\eea
 when $\lambda(1,m^2_\ell,n^2_\ell)< 0$. On the other hand, when $\lambda(1,m^2_\ell,n^2_\ell) \geq
 0$ we have
\bea B_0(\ell^2,m^2,n^2)&=& D
-\ln\big(\frac{m^2}{\mu^2}\big)+2-\sqrt{\lambda(1,m^2_\ell,n^2_\ell)}\bigg\{\ln\bigg[\sqrt{\frac{1}{4
m^2_\ell}}\big(1+m^2_\ell-n^2_\ell\big)+\sqrt{\frac{1}{4
m^2_\ell}\lambda(1,m^2_\ell,n^2_\ell)}\bigg]\nonumber\\
&+& \ln\bigg[\sqrt{\frac{1}{4
n^2_\ell}}\big(1+n^2_\ell-m^2_\ell\big)+\sqrt{\frac{1}{4
n^2_\ell}\lambda(1,n^2_\ell,m^2_\ell)}\bigg]-i\pi\bigg\}+\frac{1}{2}\big(1+n^2_\ell-
m^2_\ell\big)\ln\big(\frac{n^2_\ell}{m^2_\ell}\big),\nonumber\\\eea
For some special cases of $B_0(\ell^2,m^2,n^2)$, we find that \bea
B_0(m^2,0,n^2) &=& D
-\ln\big(\frac{n^2}{\mu^2}\big)+2+\big(\frac{n^2}{m^2}-1\big)\ln\big(1-\frac{m^2}{n^2}\big)
\,\,\,\, if\,\, m^2< n^2, \nonumber\\
B_0(m^2,0,n^2) &=& D
-\ln\big(\frac{n^2}{\mu^2}\big)+2+\big(\frac{n^2}{m^2}-1\big)\bigg[\ln\big(\frac{m^2}{n^2}-1\big)-i
\pi\bigg] \,\,\,\, if\,\, m^2 \geq n^2, \nonumber\\ B_0(0,m^2,n^2)
&=& D
-\ln\big(\frac{m^2}{\mu^2}\big)+1+\frac{n^2}{m^2-n^2}\ln\big(\frac{n^2}{m^2}\big),
\nonumber\\
B_0(0,m^2,n^2) &=& \frac{A_0(m^2)-A_0(n^2)}{m^2-n^2}, \nonumber\\
B_0(m^2,0,m^2)&=& 1+\frac{A_0(m^2)}{m^2}. \eea Finally, the loop
function $ C_0(m^2_1,\kappa^2,m^2_2,\ell^2,m^2,n^2)$ can be
defined as \bea C_0(p^2_1,(p_1+p_2)^2,p^2_2,\ell^2,m^2,n^2)&=&
\frac{16\pi^2 \mu^{4-d}}{i}\int\frac {d^d k}{(2\pi
)^d}\frac{1}{(k^2-\bar \ell^2)((k+p_1)^2-\bar
m^2)\big[(k-p)^2-\bar n^2\big]}\bigg],\nonumber\\\eea where
$p^2_1= m^2_1$, $p^2_2= m^2_2$ and $(p_1+p_2)^2= \kappa^2$. After
Feynman parameterization and  shifting the loop momentum $k$ to
absorb the terms linear in $k$  one can proceed to obtain the
final result of the integration.

\end{document}